# Weak antilocalization and zero-field electron spin splitting in AlGaN/AlN/GaN heterostructures with a polarization induced two-dimensional electron gas

Ç. Kurdak[(1),(2)], N. Biyikli[(2)], Ü. Özgür[(2)], H. Morkoç[(2)], and V. I. Litvinov[(3)]

[(1)] *Physics Department, University of Michigan, Ann Arbor, MI 48109*
[(2)] *Department of Electrical Engineering, Virginia Commonwealth University, Richmond VA 23284*
[(3)] *WaveBand/Sierra Nevada Corporation, 15245 Alton Parkway, Suite 100 Irvine, CA 92618*

Spin-orbit coupling is studied using the quantum interference corrections to conductance in AlGaN/AlN/GaN two-dimensional electron systems where the carrier density is controlled by the persistent photoconductivity effect. All the samples studied exhibit a weak antilocalization feature with a spin-orbit field of around 1.8 mT. The zero-field electron spin splitting energies extracted from the weak antilocalization measurements are found to scale linearly with the Fermi wavevector ($E_{SS} = 2\alpha k_f$) with an effective linear spin-orbit coupling parameter $\alpha = 5.5 \times 10^{-13}$ eV·m. The spin-orbit times extracted from our measurements varied from 0.74 to 8.24 ps within the carrier density range of this experiment.

PACS Numbers: 71.70.Ej, 73.20.Fz, 73.40.Kp

Following the advances in magnetic semiconductors, there has been growing interest in spin-based electronics (spintronics).[1] Realization of useful spintronic devices requires controlled spin polarization, spin transport, and spin detection. These are particularly challenging tasks. Unlike charge, the spin of an electron in a semiconductor system is a nonconserved quantity mainly due to spin-orbit coupling. Spin-orbit interaction in zincblende III-V semiconductor quantum wells manifests itself in the Dresselhaus[2] and Rashba[3] effects arising from bulk inversion asymmetry of the crystal and the structural inversion asymmetry of the confinement potential, respectively. The spin-splitting that arises from the Rashba effect is isotropic and scales linearly with the Fermi wavevector $k_F$ whereas there are two terms for spin-splitting associated with the Dresselhaus effect; one scales as $k_f$ and the other scales as $k_f^3$ but is anisotropic in the plane of the quantum well. The Rashba coupling is of particular interest for spin transistor applications, as it can be controlled by a gate potential.[4]

Both low and high bandgap semiconductors are currently being considered for spintronic applications. In low bandgap semiconductors, such as InAs, the Rashba coupling is known to be strong, which is desirable for spin transistor applications. On the other hand, dilute magnetic semiconductors based on low bandgap semiconductors exhibit low Curie temperatures above which ferromagnetism is lost. In contrast, it has been suggested that at room temperature, or even above, ferromagnetism can be achieved in high bandgap semiconductors such as GaN and ZnO.[5,6] Within this context, there has been recent interest in exploring AlGaN/GaN heterostructures for spintronic applications.

Spin-orbit interaction and the associated spin-splitting in zincblende III-V semiconductor heterostructures have been studied for more than a decade and are relatively well understood.[7] In addition to Rashba and Dresselhaus terms for spin-splitting there is an additional term for wurtzite quantum well structures that arises from bulk inversion asymmetry with a functional form identical to that of the Rashba term.[8] These terms have not been measured independently in the wurtzite system. Moreover, recent experiments based on Shubnikov-de Haas (SdH),[9,10] weak antilocalization (WAL),[11,12,13] and circular photogalvanic[14] measurements have given conflicting results for the spin splitting in wurtzite AlGaN/GaN heterostructures. In particular, spin-splitting energies extracted from the beat pattern of SdH measurements are found to be as large as 9 meV, which is about an order of magnitude larger than the theoretical estimates based on the Rashba coupling mechanism for this material system.[15] To account for the discrepancy, Lo *et al.* have proposed an additional spin splitting mechanism for wurtzite structures[16] and Tang *et al*. proposed an alternative interpretation of such data based on magneto-intersubband scattering.[17]

To help resolve these issues, we have performed WAL and SdH measurements on three $Al_xGa_{1-x}N$/AlN/GaN samples with different Al concentrations. We used the persistent photoconductivity effect to vary the carrier density of the two-dimensional electron gas (2DEG).[18] The electron spin splitting energies extracted from our weak antilocalization measurements varied from 0.3 to 0.7 meV. Consistent with such small spin splitting energies, we have not seen any beat feature in the SdH oscillations. More importantly, the measured spin splitting energies are found to scale linearly with the Fermi wavevector.

The three heterostructures used in this study were grown by metalorganic vapor phase epitaxy on c-plane sapphire substrates and consist of the following layers: a 3 μm thick GaN buffer layer, a 1 nm thick AlN interfacial layer, a 25 nm thick $Al_xGa_{1-x}N$ layer, and 3 nm of GaN cap layer where $x = 0.1$, 0.15, and 0.25 for heterostructures A, B, and C, respectively. All layers were undoped and the 2DEG is formed as a result of spontaneous and piezoelectric polarization effects just below the AlN interfacial layer. An AlN interfacial layer between the GaN and AlGaN layers was used to suppress alloy scattering.[19] 600 μm long 100 μm wide Hall bar structures were fabricated by photolithography followed by dry etching. Ti/Al/Ti/Au contacts annealed at 900 °C were then used to form ohmic contacts to the 2DEG.

Magnetoresistance and Hall measurements were performed in a variable temperature cryostat with a base temperature of 1.6 K. The samples exhibited SdH oscillations and integer quantum Hall effect at high magnetic fields. As expected, the carrier density of the sample with the highest Al fraction (Sample C) had the highest carrier density. To change the carrier density of the samples, we have illuminated the top surface of the samples through the optical access port of the cryostat via a flashlight. After each illumination the carrier density of the sample increases and does not drop to its equilibrium concentration, unless the sample is warmed up to room temperature.[20] By using the persistent photoconductivity effect we were able to vary the carrier density of the samples in a controllable manner over the ranges of $0.8$-$1.3 \times 10^{12}$ cm$^{-2}$, $1.7$-$4.9 \times 10^{12}$ cm$^{-2}$, and $3.1$-$6.7 \times 10^{12}$ cm$^{-2}$ for samples A, B, and C, respectively. Sample B had the highest electron mobility of $\mu = 20,300$ cm$^2$/V·s at a carrier concentration of $n = 4.9 \times 10^{12}$ cm$^{-2}$. Consistent with previous studies based on gated structures, at low carrier densities the electron mobility is found to be decreasing with decreasing carrier density.[21] Typical high field magnetoresistivity traces obtained from these three samples are shown in Fig. 1. From the temperature dependence of SdH oscillations, we extracted an effective electron mass of $m^* = 0.23 m_e$. We could not resolve any beat feature in the SdH oscillations even at high carrier densities where the onset of SdH is around 2 Tesla. The SdH oscillations also indicate that only a single subband of the quantum well is occupied by the 2DEG. Furthermore, at high carrier densities, we could resolve SdH oscillations corresponding to filling fractions above 100 which implies that the carrier density was uniform throughout our device even after illumination.

The absence of any beat feature in the SdH oscillations prevented us from extracting spin-splitting energies from high magnetic field measurements. An alternative method, however, was employed to extract the spin-orbit coupling and the associated spin-splitting energies from the measurements of quantum corrections to conductance. Hikami *et al.* first predicted that in the limit of large spin-orbit interaction the quantum correction to conductance changes sign.[22] Experimentally the quantum interference corrections are typically studied by performing high accuracy magnetoconductance measurements at low magnetic fields where the sign of the magnetoconductance is either positive or negative depending on the size of spin-orbit coupling. If a negative magnetoconductance near zero magnetic field, also known as the WAL feature, is observed, then it can be concluded that the quantum correction arises from the interference of spin-dephased paths, in which case the a spin-orbit coupling parameter can be extracted from such a measurement.

To this end, we have performed magnetoconductance measurements at low magnetic fields. The low-field magnetoconductivity after the subtraction of the zero field background, $\Delta\sigma = \sigma(B) - \sigma(0)$, of sample B is shown in Fig. 2 for two different temperatures. There is a clear negative magnetoconductance behavior at magnetic fields below 2 mT. The size of WAL peak is strongly temperature dependent and decreases with increasing temperatures. At this carrier density, the WAL feature was smeared out at temperatures above 4.2 K. We have also performed measurements in tilted magnetic fields and found that the magnetoconductivity is caused by the component of magnetic field which is perpendicular to the 2DEG plane only. Qualitatively, these results are consistent with the recent WAL measurements performed on a similar AlGaN/GaN heterostructure with a 2DEG.[11-13] However, the size of the WAL feature Thillosen *et al.* reported was as large as $\Delta\sigma = 30(e^2/2\pi h)$ which is more than an order of magnitude larger than the largest WAL we have observed.[11] Note that the sizes of quantum interference corrections to conductivity are typically on the order of $e^2/2\pi h$.

The theoretical equations for the WAL correction of a 2DEG are rather complex.[23] Typically, to extract spin-orbit parameters, depending on the electron mobility of the sample, different equations corresponding to diffusive[24] or ballistic[25] regimes can be used. In this work, we used the magnetoconductance equations first calculated by Iordanskii, Lyanda-Geller, and Pikus (ILP) for the diffusive limit given in Ref. 20. The starting Hamiltonian of the ILP theory includes both isotropic and anisotropic spin splitting terms. The use of the isotropic term was found to be sufficient for our data analysis.

For the data shown in Fig. 2, we use the measured values of carrier density $n = 1.73 \times 10^{12}$ cm$^{-2}$ and mobility $\mu = 10,000$ cm$^2$/V·s from which we determine the elastic scattering rate $\tau = 1.31$ ps, the diffusion constant $D = 179$ cm$^2$/s, and the characteristic transport field $B_{tr} = \hbar/4eD\tau_{tr} = 7.02$ mT for this sample. We fit the data with two adjustable parameters, the spin-orbit field $B_{SO} = \hbar/4eD\tau_{SO}$ and the phase coherence field $B_\phi = \hbar/4eD\tau_\phi$, where $\tau_{SO}$ and $\tau_\phi$ are the spin-orbit and phase coherence times, respectively. The $\tau_{SO}$ and $\tau_\phi$ extracted from such fits are shown as a function of temperature in the inset of Fig. 2. As expected the measured $\tau_{SO}$ is found to be nearly independent of temperature whereas the $\tau_\phi$ exhibit a strong temperature dependence. The magnitude of the phase coherence time is similar to that of other III-V semiconductors measured in the same temperature range.[13, 26] In the limited temperature range of these measurements, the phase coherence rate scales $1/\tau_\phi = T^{1.35}$. Theories based on electron-electron interactions predict the phase coherence rate to scale linearly and quadratically with temperature at low and high temperatures, respectively.[27,28]

In Fig. 3 we show the low field magnetoconductivity data from the three samples at different carrier densities. There are two striking features in this data set. First, the size of the WAL peak decreases with decreasing carrier density; in fact, we could not observe the WAL feature for carrier densities below $1\times10^{12}$ cm$^{-2}$ at 1.6 K. Second, the width of the WAL appears to be independent of the carrier density. Indeed, $B_{SO}$ extracted from the WAL features is found to be around 1.8 mT for all three samples. This is the central finding of this work. Note that $B_{SO} = (\hbar/4eD)2\Omega^2\tau$, where $\Omega$ is the spin-orbit frequency. Using this equation, we calculated the isotropic spin-splitting energy $E_{SS} = 2\hbar\Omega$ and plotted it as a function of Fermi wavevector $k_f$ in Fig. 4. There is a clear overlap between the data extracted from samples B and C indicating the spin splitting energies do not directly depend on the Al composition. It is clear that the spin-splitting energy scales linearly with $k_f$. Both the Rashba term and the bulk inversion asymmetry term unique to wurtzite systems can lead to the linear scaling of spin splitting energy. By fitting the data to a linear form $E_{SS} = 2\alpha k_f$, we extract an effective linear spin-orbit coupling parameter $\alpha = 5.5\times10^{-13}$ eV·m. This effective spin-orbit coupling parameter that we measured should be viewed as a sum of the Rashba parameter $\alpha_R$ and a coupling parameter associated with the bulk inversion asymmetry in wurtzite quantum wells $\alpha_{BIA}$. For comparison, we note that in two-dimensional systems such as GaAs/AlGaAs heterostructures with a high carrier density 2DEG, the measured spin-splitting energy is known to be proportional to $k_f^3$ and the spin-orbit interaction is mainly caused by the Dresselhaus effect.[7] In the inset of Fig. 4, we also show the $\tau_{SO}$ extracted from Sample B as a function of carrier density. Sample A and C had the longest and shortest spin-orbit times of $\tau_{SO} = 8.24$ ps and $\tau_{SO} = 0.74$ ps with elastic scattering rates of $\tau = 1.02$ ps and $\tau = 2.33$ ps at carrier densities of $n = 1.32 \times 10^{12}$ cm$^{-2}$ and $n = 6.74 \times 10^{12}$ cm$^{-2}$, respectively.

In summary, we have studied spin-orbit coupling in AlGaN/AlN/GaN two dimensional electron systems using magnetotransport measurements. At low magnetic fields we have observed WAL behavior from which we extracted the spin-orbit and phase coherence times. The spin splitting energies extracted from WAL measurements ranged from 0.3 to 0.7 meV and were found to scale linearly with $k_f$. The effective linear spin-orbit coupling parameter $\alpha = 5.5 \times 10^{-13}$ eV·m that we determined from our measurements is significantly smaller than previous reports based on SdH measurements.

We would like to thank V. Avrutin for useful discussions. This work is supported by grants from the Air Force Office of Scientific Research (AFOSR) under the direction of Dr. G. L. Witt and Dr. K. Reinhart and also by the Missile Defense Agency contract W9113M-04-C-0088 to WaveBand Corporation.

# FIGURES

**Fig. 1.** (Color Online) Typical longitudinal resistivity versus magnetic field traces for three samples with a wide range carrier densities at 1.6 K. From top to bottom, the first two traces are for sample A, next two curves are for sample B, and the lowest three curves are obtained from sample C. The corresponding carrier densities for the top and bottom traces are $n = 1.1 \times 10^{12}$ cm$^{-2}$ and $n = 6.7 \times 10^{12}$ cm$^{-2}$, respectively. The persistent photoconductivity effect is used to vary the carrier density of each sample.

**Fig. 2.** (Color Online) Experimental magnetoconductivity $\Delta\sigma = \sigma(B) - \sigma(0)$ of Sample B at a carrier density of $n = 1.73 \times 10^{12}$ cm$^{-2}$ in units of $e^2/2\pi h$ at 1.8 K (circles) and 3 K (triangles). The solid lines are theoretical fits to the data. The inset shows $\tau_{SO}$ (circles) and $\tau_\phi$ (squares) extracted from the theoretical fits as a function of temperature.

**Fig. 3.** (Color Online) Experimental magnetoconductivity $\Delta\sigma = \sigma(B) - \sigma(0)$ of the three samples at different carrier densities measured near the base temperature of our cryostat (1.6 to 1.8 K). The solid lines are theoretical fits to the data.

**Fig. 4.** (Color Online) The isotropic spin-splitting energy $2\hbar\Delta$ extracted from the WAL measurements versus Fermi wavevector. The solid line is a linear fit to the data. The inset shows $\tau_{SO}$ of sample B as a function of carrier density.

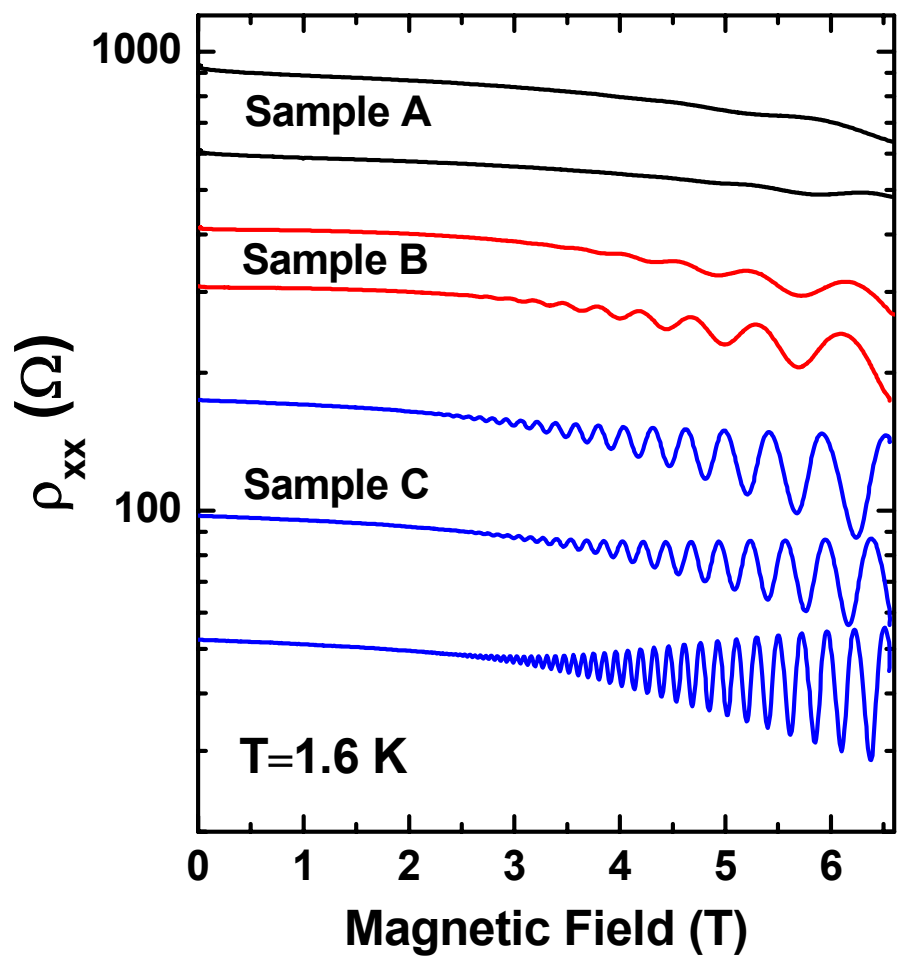

Figure 1

Kurdak *et al*.

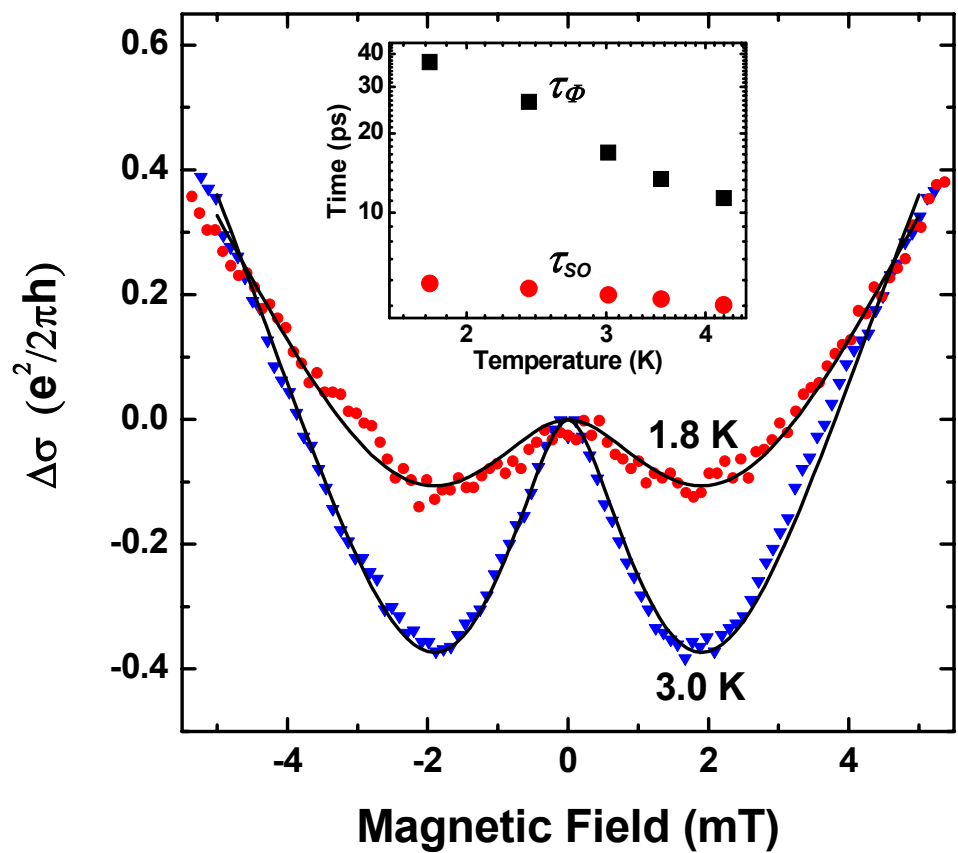

Figure 2

Kurdak *et al*.

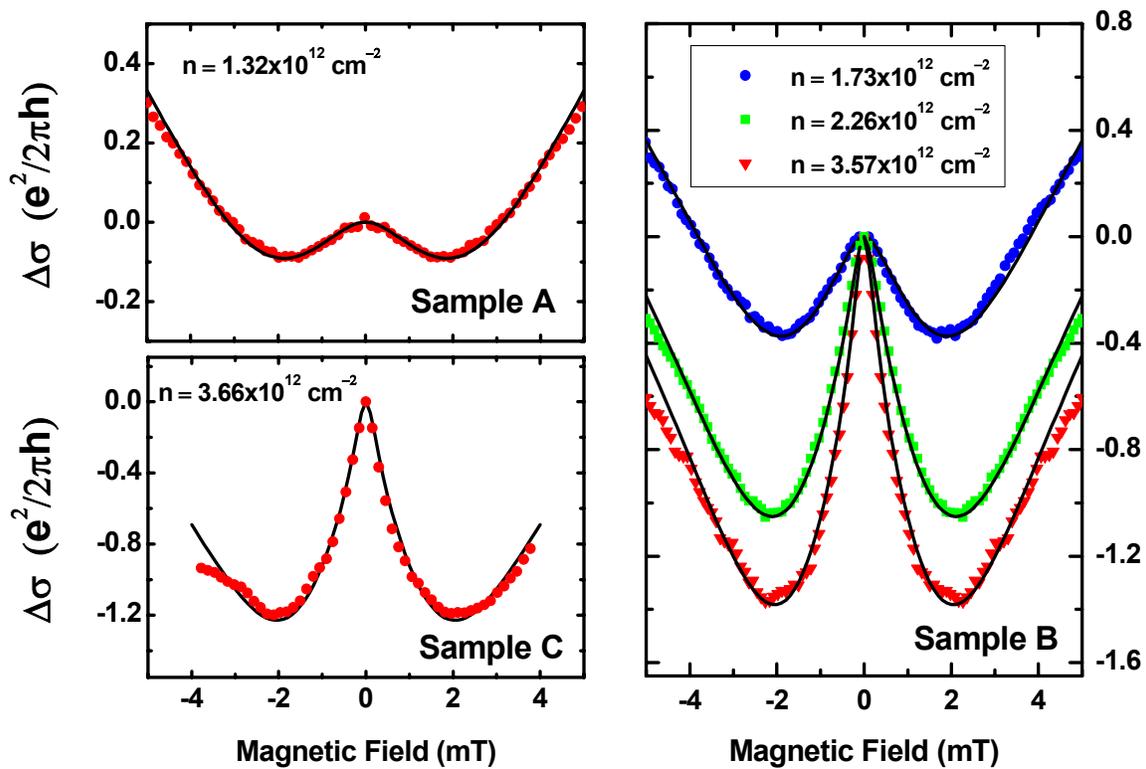

Figure 3

Kurdak *et al*.

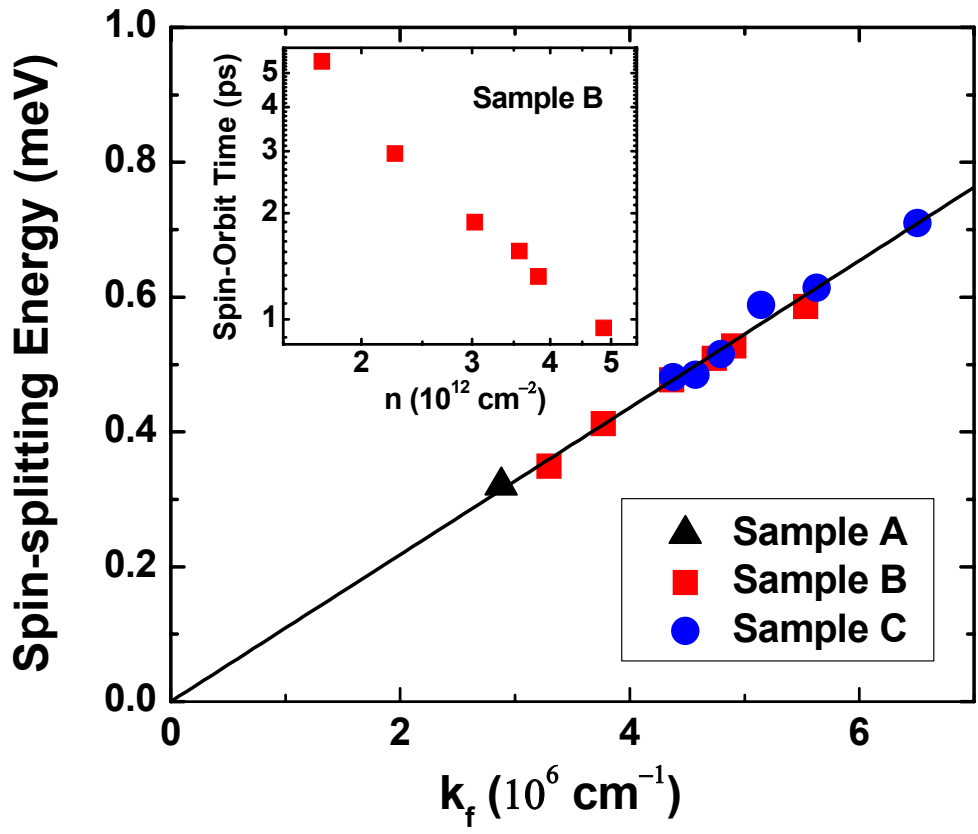

Figure 4

Kurdak *et al*.